\documentclass[twocolumn,showpacs,showkeys,preprintnumbers,amsmath,amssymb]{revtex4}

\usepackage{epsfig}
\usepackage{amsfonts}
\usepackage{float}
\usepackage{dcolumn}
\usepackage{bm}

\floatstyle{ruled}
\newfloat{algorithm}{htbp}{loa}
\floatname{algorithm}{Algorithm}  

\begin{document}
\title{Multiplexing of discrete chaotic signals in presence of noise}

\author{Nithin Nagaraj}%
 \email{nithin_nagaraj@yahoo.com}\homepage{http://nithin.nagaraj.googlepages.com}
  \author{Prabhakar G. Vaidya}
\affiliation{%
School of Natural Sciences and Engineering, National Institute of
Advanced Studies, Indian Institute of Science Campus, Bangalore 560
012.
}%

\date{\today}

\begin{abstract}
In this paper, multiplexing of discrete chaotic signals in the
presence of noise is investigated. Existing methods are based on
chaotic synchronization which is susceptible to noise and parameter
mismatch. Furthermore, these methods fail for multiplexing more than
two discrete chaotic signals. We propose two novel methods to
multiplex multiple discrete chaotic signals based on the principle
of symbolic sequence invariance in the presence of noise and finite
precision implementation of finding the initial condition of an
arbitrarily long symbolic sequence of a chaotic map.
\end{abstract}

\pacs{05.45.Gg}%
 \keywords{Multiplexing, chaos communication, discrete chaotic maps, lossless compression.}

\maketitle
\section{Motivation}
In this paper, we investigate multiplexing of chaotic signals in the
presence of noise. The fact that chaotic signals have a lot of
redundancy in them is exploited here. We first motivate the
application and then provide new methods.

Multiplexing of signals is a very important requirement in most
communication systems. Consider the scenario where there are
multiple signals from multiple senders to be transmitted to multiple
receivers, but there exists only one communication channel that can
transmit only one signal at any given time. In such a scenario, it
would be beneficial if all the signals are ``added'' in a special
way to create a single composite signal for transmission across the
communication channel. This single composite signal is ``separated''
to the respective signals in a lossless fashion at the other end of
the channel. Note that there is also noise which is invariably added
at the channel. This scenario can also occur in transmission of
neuronal signals from different parts of the brain to various parts
of the body through a single channel.

For linear communication systems, standard ways such as frequency
division multiplexing (different signals are allocated different
parts of the frequency spectrum) and time division multiplexing
(different signals are allocated different time slots for
transmission) are used to increase the information capacity of the
channel~\cite{GibsonBook}. Nonlinear chaotic oscillators are
increasingly being used in communications since it offers a
potential advantage over conventional classical methods in terms of
noise performance~\cite{Multiplex1}. Multi-user chaotic
communications has become a hot topic of research in recent
times~\cite{Multiplex2}. It is also potentially useful in
spectrum-spreading communication systems. Hence there is a need for
multiplexing chaotic signals.

\section{Existing work and their limitations} There has been some
work in multiplexing chaotic signals. For the first time in 1996,
multiplexing of chaos using chaotic synchronization was investigated
in a simple map and an electronic circuit model by Tsimring and
Sushchik~\cite{Multiplex3}. Liu and Davis~\cite{Multiplex4} used a
scalar signal to simultaneously synchronize two different pairs of
chaotic oscillators. They called this method as dual
synchronization. However, there are several limitations of this
method. They derive a condition for dual synchronization which holds
only for certain discrete chaotic signals (maps) and for certain
values of the coupling coefficients. The notable omission is the
binary map (Bernoulli shift). They show that the binary map does not
satisfy the condition for dual synchronization for any value of the
coupling coefficients. Thus, chaotic signals from the binary map
can't be multiplexed by their method. Another limitation is that
their method can only work with two chaotic signals. It is not known
whether the method can be extended to multiple signals (more than
two) from different maps.

There has been more work in multiplexing chaotic signals from
continuous chaotic systems (flows). Liu and Davis~\cite{Multiplex4}
extend their work to multiplex signals from delay-differential
equations. Further progress has been made by Di Ning et.
al.~\cite{Multiplex5} who extend Liu's method for 3D continuous
chaotic systems (Lorenz and R\"{o}ssler systems). This has been
further improved by Salarieh and Shahrokhi~\cite{Multiplex6} who
make use of a time-varying output feedback strategy to achieve dual
synchronization without the need for all the master states (only a
linear combination of the master states are enough).

All the developments were only for two chaotic systems until the
work of Salarieh and Shahrokhi~\cite{Multiplex7} who succeeded in
multiplexing more than two continuous chaotic signals using chaotic
synchronization via output feedback strategy. They derive a
necessary condition for multi-synchronization and demonstrate the
algorithm to the Chen-Lorenz-Rossler and the Duffing-Van der Pol
continuous time chaotic dynamical systems. However, none of these
methods work for multiple {\it discrete} chaotic signals. Another
serious limitation is that chaotic synchronization is susceptible to
noise and to parameter mismatch. Even one percent of parameter
mismatch leads to $8\%$ of synchronization error and one percent of
noise results in a synchronization error of $4\%$ as reported by
\cite{Multiplex4}.

Vaidya's method~\cite{Vaidya-noise-restent} which multiplexes more
than two discrete chaotic signals remains as the latest development
on multiplexing chaotic signals from discrete chaotic signals (1D
maps). Vaidya's method does not use chaotic synchronization and is
fundamentally different from the previous approaches. We will
describe the method and its drawbacks in
Section~\ref{subsection:vaidyamethoddrawbacks} since we are going to
use some of the ideas from this method to improve upon it. In
principle, it is possible to extend the methods that is proposed in
this paper to flows and to higher dimensional chaotic dynamical
systems, but these will not be pursued here. Since Liu and Davis'
method does not work for the standard binary map, we shall consider
chaotic signals from the standard binary map with randomly chosen
initial conditions.

\subsection{New Approach}
\begin{figure}[!h]
\centering
\includegraphics[scale=.4]{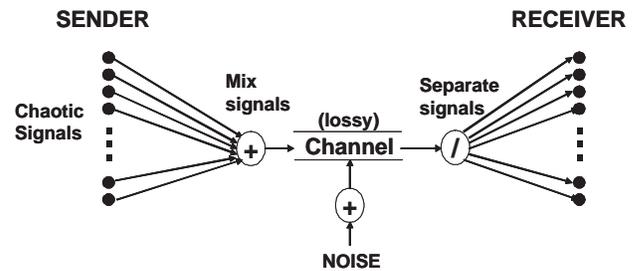}\\
(a) Scenario 1: Channel is lossy, noise is additive and limited in
magnitude
(Methods 1 and 2).\\
\vspace{0.1in}
\includegraphics[scale=.4]{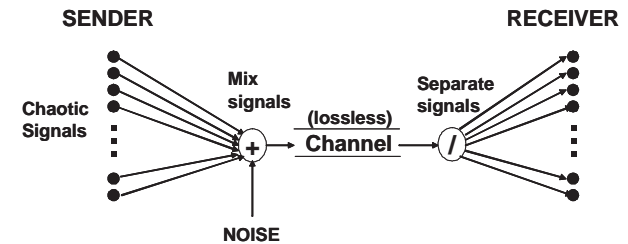}\\
(b) Scenario 2: Channel is lossless, but noise which has the same
magnitude as the signal, is added in a special way at the sender
(Method 3). \caption[Multiplexing of chaotic signals in the presence
of noise]{Multiplexing of chaotic signals in the presence of noise.}
\label{fig:figmultiplex1}
\end{figure}

Our approach considers two different scenarios as shown in
Figure~\ref{fig:figmultiplex1}. In Scenario 1, the communicating
channel is noisy and there is no control on the noise that is added
during transmission of the signal
(Figure~\ref{fig:figmultiplex1}(a)). However, it shall be assumed
that the magnitude of the noise is limited (we shall give conditions
on the magnitude of noise that is allowed by our methods). The noise
is uniformly distributed. The signals are chaotic and the noise that
is added during transmission is uncontrolled but limited in
magnitude.

In Scenario 2 (Figure~\ref{fig:figmultiplex1}(b)), the communication
channel is lossless, but the noise is added at the sender. The noise
is assumed to be of the same magnitude as the chaotic signals and
uniformly distributed. But the way the noise is added is under
control. This scenario corresponds to
steganography~\footnote{Steganography is the art and science of
hiding secret data in a message. Unlike cryptography where it is
known that the data is encrypted and meant to be secret, in
steganography, the presence of the secret information itself is
hidden.} or cryptographic applications where the noise could be the
``payload'' to be secretly transmitted.

Recently, Vaidya~\cite{Vaidya-noise-restent} suggested a novel
multiplexing algorithm for 1D discrete chaotic signals in the
presence of noise. We shall call this as Method 1 and review it
briefly and list some of its limitations. A new method (Method 2)
will be proposed which overcomes some of the limitations of Method
1. Both these methods are solutions to Scenario 1. A novel method
(Method 3) is proposed for Scenario 2.

\section{Method 1: Vaidya's Noise Resistant Map} Vaidya's method~\cite{Vaidya-noise-restent} is a
solution to multiplexing of chaotic signals in the presence of noise
(Scenario 1) which does not make use of chaotic synchronization like
the previous approaches. Vaidya proposes a noise-resistant version
of the Tent map. Since we are going to deal mainly with the standard
binary map, a minor modification leads to a noise-resistant version
of the binary map. It is given by the following set of equations:

\begin{eqnarray*}
y & = & 2x,~~~~~~~~~~~~ 0 \leq x < \frac{p}{2}\\
  & = & 2x+q,~~~~~~~~~~~~ \frac{p}{2} \leq x < p\\
  & = & 0,~~~~~~~~~~~~~~~ p \leq x < p+q\\
  & = & 2x - 2(p+q),~~~~~~ p+q \leq x < \frac{3p}{2}+q\\
  & = & 2x - (2p+q),~~~~~~ \frac{3p}{2}+q \leq x < 2p+q \\
  & = & 0,~~~~~~~~~~~~~~~  2p+q \leq x < 1.
\end{eqnarray*}

\begin{figure}[!h]
\centering
\includegraphics[scale=.4]{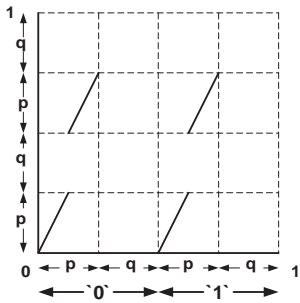}
\caption[Method 1: Noise-resistant binary map]{$T_{noiseres}$:
Noise-resistant Binary Map.} \label{fig:fignoiseresmap}
\end{figure}

Figure~\ref{fig:fignoiseresmap} depicts the noise resistant binary
map (denoted by $T_{noiseres}$). Vaidya establishes that there
exists a conjugacy between the ordinary binary map and
$T_{noiseres}$. Given any chaotic signal on the binary map (a
chaotic signal is trajectory on the map for a given initial
condition), one can find the equivalent signal on $T_{noiseres}$.
$p$ ($0<p \leq 0.5$) and $q$ ($0\leq q < 0.5$) can be chosen such
that $2p+2q=1$.

The symbolic sequence is defined as follows:
\begin{eqnarray*}
 S(x_i) &=& 0,~~~~~~ 0 \leq x_i < p+q\\
        &=& 1,~~~~~~ p+q \leq x_i < 1.
\end{eqnarray*}
Here, $X = \{ x_i \}_{i=1}^{i=m}$ is the chaotic trajectory (or
chaotic signal) starting from an initial condition $x_1$. $S(X)$
denotes the symbolic sequence for the entire trajectory.

\subsection{Noise-resistance}
For any given chaotic signal $X$ on $T_{noiseres}$, if noise $N = \{
n_i \}_{i=1}^{i=m}$ is added such that each $n_i$ satisfies $0 \leq
n_i < q$, then it can be seen that the symbolic sequence remains
unchanged:

\begin{equation}
S(X) = S(X+N).
\end{equation}

The signal $X$ is transmitted at the sender and the resulting signal
$Z = X+N$ is what is received at the receiver. However, because of
the above property of {\it symbolic sequence invariance}, we can
compute $S(X)$ (=$S(Z)$) and iterating backwards on the map, we can
find the initial condition $x_1$. Knowing $x_1$, we can easily
compute $x_2, \ldots, x_{m}$ and thus we can recover $X$ losslessly.
Furthermore, we can also compute $N = Z-X$ accurately. Thus we have
recovered both the original chaotic signal and the noise. This noise
resistance property is provided by a non-zero value of $q$. The
larger the $q$, the higher is the resistance to noise but at the
same time the length of the signal has to be longer in order to
determine the initial condition $x_1$ accurately from the symbolic
sequence.

\subsection{Cascading Noise-resistant Maps}
Vaidya goes one step further and defines a cascade of such
noise-resistant maps. To add another chaotic signal $Y$ to $X$,
Vaidya defines a similar noise-resistant map which maps $[0,q)$ onto
itself. It is self-similar to $T_{noiseres}$. Thus, he defines a
whole cascade of noise-resistant maps, all of which are self similar
to the original one. The domain of succeeding maps reduces
exponentially. For further details, please refer to
\cite{Vaidya-noise-restent}.

With these cascade of maps, one could now add a whole family of
chaotic signals $\{ X_1, X_2, \ldots, X_k \}$ to one noise signal
$N$ (on the channel) with magnitude dictated by the number of maps,
to yield the signal $Z$. The symbolic sequence invariance is
maintained at each step of cascading. Thus, the symbolic sequence of
$Z$ is used to decode $X_1$ and $N_1$ where $N_1$ is the sum of $\{
X_2, X_3, \ldots, X_k \}$ and $N$. The symbolic sequence of $N_1$ is
the same as that of $X_2$ and hence $X_2$ can be decoded. This
procedure is repeated until all the signals are losslessly recovered
along with $N$.  Vaidya successfully applies this method to
multiplex 100 chaotic signals.

\subsection{Drawbacks of Method 1}
\label{subsection:vaidyamethoddrawbacks}

The drawbacks of Method 1 are as follows:

\begin{enumerate}
\item Given chaotic signals from various chaotic maps, one has to find
the corresponding signals in the noise-resistant binary/tent map
using topological conjugacy. This is cumbersome and although we can
have a finite precision implementation of the chaotic maps and the
noise-resistant binary map, it is difficult to develop finite
precision implementation of the topological conjugacy. For example,
if the signal $X$ is from the logistic map, then the topological
conjugacy will involve trigonometric functions which would have to
be expressed as truncated infinite series. This may lead to errors
in the recovery of $X$.

\item The amount of noise that can be added reduces exponentially as the
number of chaotic signals to be multiplexed increases linearly.

\item It is in principle possible to extend the idea of noise
resistant maps to other chaotic maps like the logistic map. However,
for each new map, the equations have to be worked out explicitly.
\end{enumerate}

We are motivated to invent new methods of multiplexing which will
circumvent the above problems. While exploiting the idea of symbolic
sequence invariance under the addition of noise, we would like to
devise a method which will work for any 1D chaotic unimodal map (and
generalizable to other kinds of maps and higher dimensional ones)
without the necessity of topological conjugacy. The scenario where
the magnitude of noise is equal to that of the signal also needs to
be addressed.

\section{Method 2} The key idea of Method 1 is the notion of
symbolic sequence invariance. As long as we ensure that the symbolic
sequence of the original chaotic signal $X$ is unaffected by adding
the noise $N$ (uniformly distributed), the resulting signal $Z=X+N$
has the same symbolic sequence as $X$ ($S(Z) = S(X)$). Then, given
this arbitrarily long symbolic sequence $S(X)$, the problem reduces
to determining its initial condition and iterating this initial
condition to obtain the entire chaotic signal $X$. Once $X$ is
determined, one could subtract $X$ from $Z$ to obtain the noise
signal $N$.

In order to find the initial condition from an arbitrarily long
symbolic sequence of a chaotic map, we make use of GLS-coding.
Generalized Lur\"{o}th Series or GLS-coding for short, is a new
entropy coding algorithm~\cite{NithinGLS} that achieves the
Shannon's entropy rate for lossless data compression. The idea of
GLS-coding is to first embed the stochastic (binary) i.i.d source
into the appropriate GLS and then treating the message as a symbolic
sequence, the initial condition is determined by a backward
iteration. A finite precision implementation of GLS-coding is
described in the Appendix. Using such an implementation, it is
possible to determine accurately the initial condition of an
arbitrarily long symbolic sequence. Our implementation is for the
skew-tent and skew-binary maps and can be extended to other maps.
The standard tent map and binary map are part of this family.

The algorithm for Method 2 is described as follows:

\begin{enumerate}
\item Let $X_1, X_2, \ldots X_k $ be $k$ chaotic signals of length $m$ to be
multiplexed. Each of these signals are obtained from distinct
initial conditions (randomly chosen) on the standard binary map.

\item Compute the symbolic sequence $ \{ SX_i \}_{i=1}^{i=k} = \{ S(X_i(1)),
S(X_i(2)), \ldots S(X_i(m)) \}_{i=1}^{i=k}$. The function S(.) is
defined as follows:

\begin{eqnarray}
 S(x) &=& 0,~~~~~~ 0 \leq x < 0.5  \nonumber\\
        &=& 1,~~~~~~ 0.5 \leq x < 1.\label{eqn:symbolicseq}
\end{eqnarray}

\item Compute $\{ S_i \}_{i=1}^{i=m}$ $=$ $\{ <$ $SX_1(1)$$SX_2(1)$$\ldots$ $SX_k(1)$$>_2$, $<SX_1(2)SX_2(2)$$\ldots SX_k(2)>_2$  $\ldots <SX_1(m)SX_2(m)$$\ldots
SX_k(m)>_2\}$. Here, $<.>_B$ denotes a number that is base-$B$
representation.

\item Compute $ D = \{ D_i \}_{i=1}^{i=m}$ where $D_i = <S_i>_{10}$.

\item Compute $Z = \{ Z_i \}_{i=1}^{i=m}$ where $Z_i =
\frac{2D_i+1}{2^{k+1}}$.

\item Transmit $Z$ across the channel.

\item Receive $Z_{noisy} = Z + N$. Here $N = \{ N_i \}_{i=1}^{i=m}$
where each $N_i$ is uniformly distributed noise in the range $(
-2^{-(k+1)}, +2^{-(k+1)} )$.

\item At receiver, compute $D_{noisy} =  \lfloor 2^{k} Z_{noisy}
\rfloor$ where $\lfloor (.) \rfloor$ is the {\it floor} operation
which computes the maximum integer that is less than the argument.
Note that $D_{noisy} = D$. This is because $D < 2^{k}Z_{noisy} <
(D+1)$. The floor operator makes this equal to $D$ since $D$ is
always a positive integer. This is where we have made use of the
fact that the symbolic sequence is invariant in spite of noise. Here
$D$ has the information of the symbolic sequence of all the $k$
chaotic signals.

\item Once we have $D_{noisy} = D$, we can recover the symbolic
sequences of each of the $k$ chaotic signals and thereby recover the
$k$ initial conditions $\{X_1(1), X_2(1), \ldots, X_k(1) \}$ by
GLS-coding (see Appendix).

\item From the initial conditions, the $k$ chaotic signals can be
recovered.

\item The noise signal can be recovered at the receiver by computing
$Z$ from $D$ and computing $N =Z_{noisy} - Z $.

\end{enumerate}

\begin{figure}[!h]
\centering
\includegraphics[scale=.5]{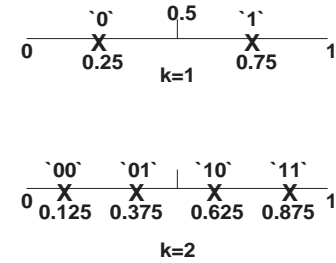}
\caption[Examples of signal points transmitted for Method 2]{Method
2: For the cases $k=1$ and $k=2$. The points marked `$X$' are
transmitted depending on the symbolic sequences of the chaotic
signals. Owing to noise at the channel, the received signal will
also be uniformly distributed in the range $[0,1)$. This explains
Figure~\ref{fig:figmethod2_Z}(c).} \label{fig:figmethod2eg}
\end{figure}

\begin{figure}[!h]
\centering
\includegraphics[scale=.5]{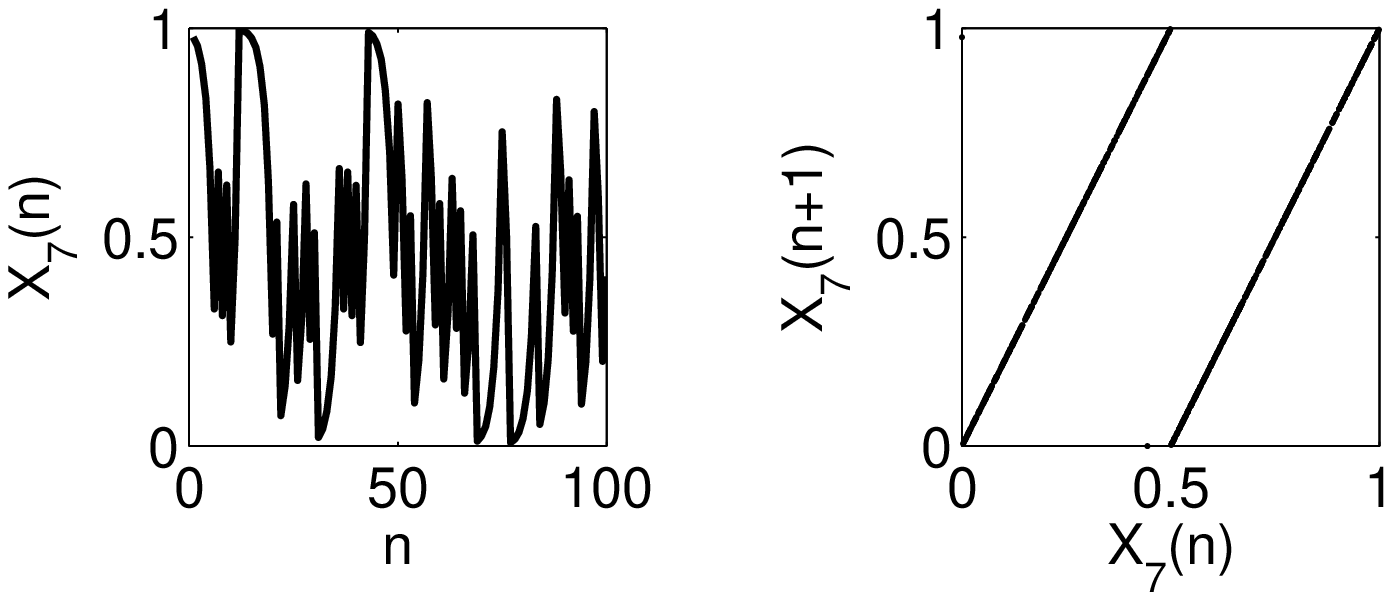}\\
\includegraphics[scale=.4]{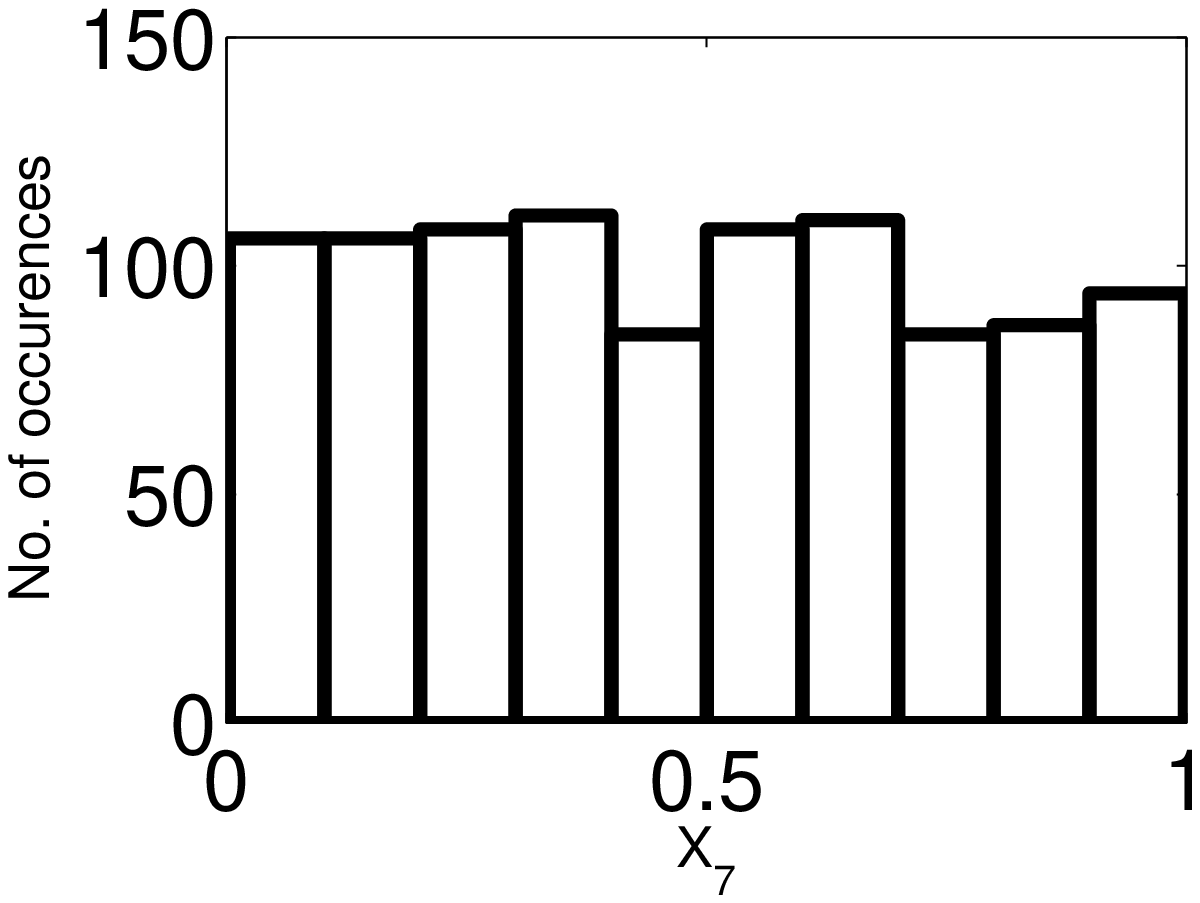}\\
\caption[Method 2: Multiplexing of chaotic signals in the presence
of noise]{Method 2: Multiplexing of $k=10$ chaotic signals in the
presence of one noise signal $N$ -- (a) Top left: The 7th Chaotic
Signal $X_7(.)$. $m=1000$, but only first 100 shown, (b) Top right:
Phase Portrait of $X_7(.)$, (c) Bottom: Histogram of $X_7(.)$.}
\label{fig:figmethod2_X7}
\end{figure}


\begin{figure}[!h]
\centering
\includegraphics[scale=.5]{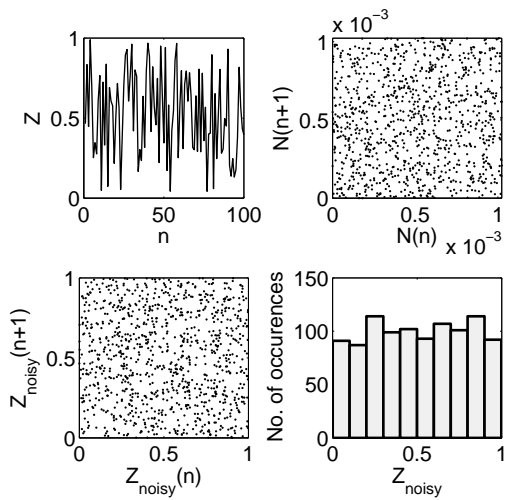}
\caption[Final signal transmitted for Method 2]{Method 2:
Multiplexing of chaotic signals in the presence of noise --(a) Top
left: $Z$ signal (showing first 100 values only),  (b) Top right:
Phase portrait of noise signal $N$, (c) Bottom left: Phase portrait
of $Z_{noisy}$, (d) Bottom right: Histogram of $Z_{noisy}$.}
\label{fig:figmethod2_Z}
\end{figure}


\subsection{Experimental Simulations} As a simple example,
Figure~\ref{fig:figmethod2eg} shows the points transmitted for the
case $k=1$ and $k=2$.

Method 2 was experimentally simulated for $k=10$ chaotic signals of
length $m=1000$ each. They were all generated by randomly chosen
initial conditions on the standard binary map.
Figure~\ref{fig:figmethod2_X7} shows the seventh chaotic signal
$X_7$. The phase portrait and the histogram are also shown.
Figure~\ref{fig:figmethod2_Z} shows $Z$, the phase portrait of noise
signal $N$, phase portrait of $Z_{noisy}$ after the addition of
noise and the histogram of $Z_{noisy}$. All the 10 chaotic signals
and the noise signal were successfully recovered in a lossless
fashion at the receiver. We have used the finite precision
implementation of GLS-coding as proposed in Appendix. This confirms
the efficacy of Method 2.

\section{Method 3} The biggest advantage of Method 2 is that in principle it
works for any dynamical system. As long as we know the Markov
partitions of the dynamical system, we can define the symbolic
sequence and hence use Method 2. There is no need of using
topological conjugacy or construction of special noise-resistant
maps like Method 1. However, one needs to develop an analog of
GLS-coding (i.e. finding initial condition for an arbitrary long
symbolic sequence of the dynamical system as given in
Algorithm~\ref{alg:FinitePrecGLS} in Appendix) for the method to
work.

Method 2 works in the presence of a lossy channel, the noise being
additive, but the magnitude of noise that is tolerated depends on
the number of signals being multiplexed. As the number of signals
($k$) increases, the magnitude of noise ($2^{-k}$) that can be
tolerated at the channel, goes down exponentially. Method 3
overcomes this limitation. The noise magnitude can be equal to the
signal. However, we can no longer operate in Scenario 1. We assume
that we have control on the ``way'' the noise is added (noise is
still assumed to be uniformly distributed) and that the channel is
lossless (Scenario 2).

Method 3 is described as follows:

\begin{enumerate}
\item Let $X_1, X_2, \ldots X_k $ be $k$ chaotic signals of length $m$ to be
multiplexed. Each of these signals are obtained from distinct
initial conditions on the standard binary map.

\item Let noise signal be $N = \{ N_i \}_{i=1}^{i=m}$
where each $N_i$ is independent and identically distributed
(uniform) in the range $(0,1)$. Noise signal $N$ is independently
generated but available at the receiver.

\item Given two signals $A = \{ A(i) \}_{i=1}^{i=m}$ and $B = \{ B(i) \}_{i=1}^{i=m}$ where $A$ is a chaotic signal
($B$ can be anything), we define the operation $A + B.S(A) = \{ A(i)
+ B(i).S(A(i)) \}_{i=1}^{i=m}$ as follows:

\begin{eqnarray*}
A(i) + B(i).S(A(i)) &=& A(i) - B(i),~~~\textrm{~if~} S(A(i))=0\\
                   &=& A(i) + B(i),~~~\textrm{~if~} S(A(i))=1.
\end{eqnarray*}
where $S(.)$ is defined in Equation~\ref{eqn:symbolicseq}.

\item Compute the following signals:
\begin{eqnarray*}
Z_1 & = & \frac{X_1 + N.S(X_1) + 1}{3}\\
Z_2 & = & \frac{X_2 + Z_1.S(X_2) + 1}{3}\\
Z_3 & = & \frac{X_3 + Z_2.S(X_3) + 1}{3}\\
&\vdots & \\%
Z_k & = & \frac{X_k + Z_{k-1}.S(X_k) + 1}{3}\\
Z & = & Z_k.
\end{eqnarray*}

\item Transmit $Z$ on the lossless channel. Note that the dynamic
range of $Z$ is $[0,1)$.

\item Receiver receives $Z$. By symbolic sequence invariance, we
have the following identities:
\begin{eqnarray*}
S(Z) = S(Z_k) & = & S(X_k)\\
       S(Z_{k-1}) & = & S(X_{k-1})\\
       & \vdots & \\ %
       S(Z_1) & = & S(X_1).
\end{eqnarray*}

\item We start with $Z$ and compute $S(Z_k)$. By the first identity,
we have $S(X_k)$. GLS-coding is applied to determine $X_k(1)$. Hence
$X_k$ is recovered losslessly. Knowing $X_k$ and $S(X_k)$, we can
compute $Z_{k-1}$ by the following equations:

\begin{eqnarray*}
Z_{k-1}(i) &=& X_k(i) -3Z(i) + 1,~~~\textrm{~if~} S(X_k(i))=0\\
           &=& 3Z(i) - X_k(i) - 1,~~~\textrm{~if~} S(X_k(i))=1.
\end{eqnarray*}

\item Knowing $Z_{k-1}$, we repeat the procedure to extract
$X_{k-1}$ and $Z_{k-2}$. This is repeated until we have extracted
all the chaotic signals and the noise signal $N$. Note that noise
can be thought of as $Z_0$ and the same procedure applies.

\end{enumerate}


\begin{figure}[!h]
\centering
\includegraphics[scale=.5]{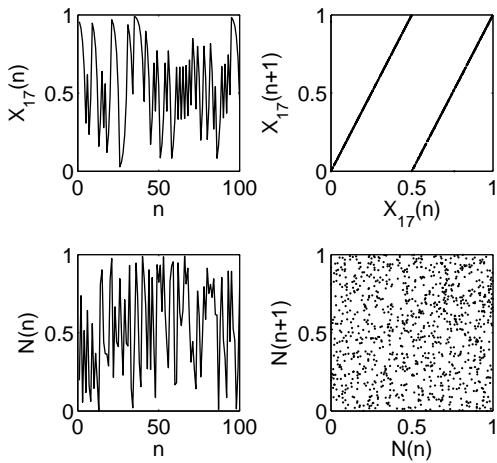}
\caption[Method 3: Multiplexing of chaotic signals in the presence
of noise]{Method 3: Multiplexing of $k=24$ chaotic signals in the
presence of one noise signal -- (a) Top left: The 17th Chaotic
Signal $X_{17}(.)$. The length $m = 1000$, but only the first 100
shown, (b) Top right: Phase Portrait of $X_{17}(.)$, (c) Bottom
left: Noise signal $N(.)$, (d) Bottom right: Phase Portrait of
$N(.)$. Note that magnitude of $N$ is the same as that of $X$.\\}
\label{fig:figmethod3_X18}
\end{figure}

\subsection{Experimental Simulations} 
\begin{figure}[!h]
\centering
\includegraphics[scale=.4]{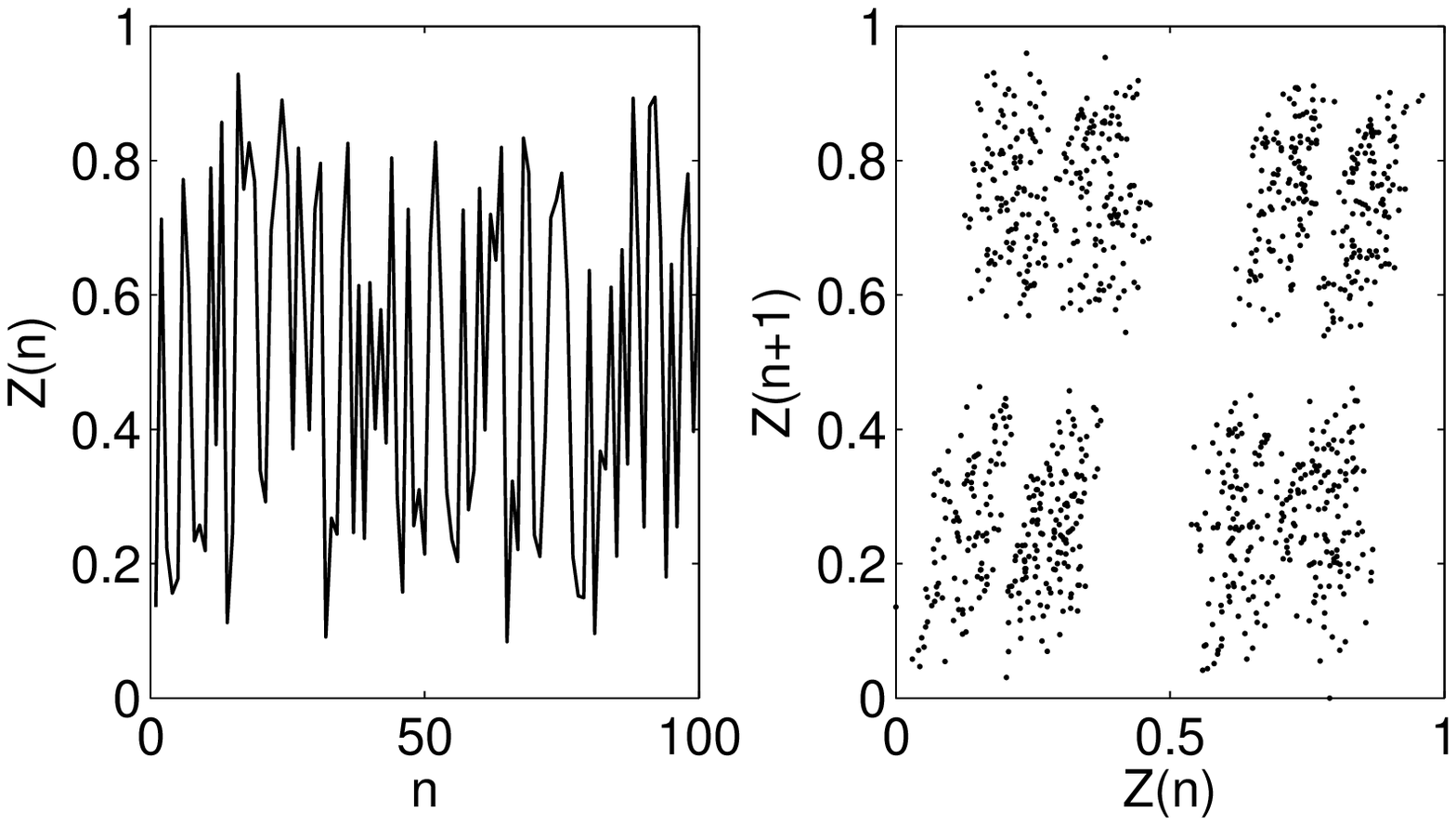}\\
\includegraphics[scale=.4]{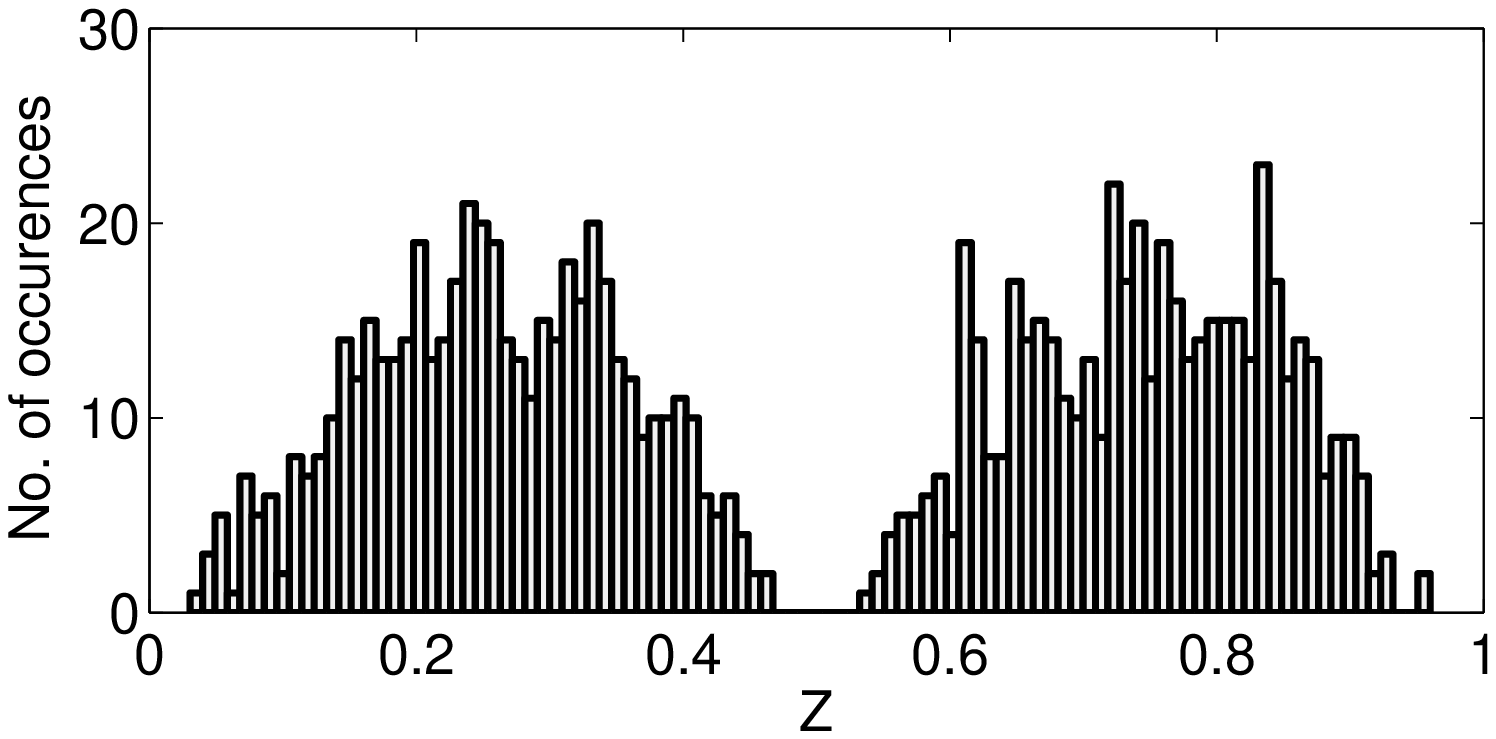}\\

\caption[Final signal transmitted for Method 3]{Method 3:
Multiplexing of chaotic signals in the presence of noise -- (a) Top
left: $Z$ signal generated by Method 3 is not uniformly distributed.
This is transmitted on the lossless channel, (b) Top right: Phase
portrait of $Z$, (c) Bottom: Histogram of $Z$ showing that it is
bimodal.} \label{fig:figmethod3_Z}
\end{figure}

Method 3 was experimentally
simulated for $k=24$ chaotic signals of length $m=1000$ each and one
noise signal of the same length. The chaotic signals were all
generated by randomly chosen initial conditions on the standard
binary map. Figure~\ref{fig:figmethod3_X18}(a) shows the 17th
chaotic signal $X_{17}$. The phase portrait is shown in
Figure~\ref{fig:figmethod3_X18}(b).
Figure~\ref{fig:figmethod3_X18}(c) shows the noise signal $N$ which
has the same magnitude as that of the chaotic signal. The phase
portrait of noise signal $N$ is shown in
Figure~\ref{fig:figmethod3_X18}(d). The final signal that is
transmitted on the lossless channel $Z$ is shown in
Figure~\ref{fig:figmethod3_Z}(a). Its phase portrait and histogram
are shown in Figure~\ref{fig:figmethod3_Z}(b) and (c) respectively.
Note that in this method, the final signal $Z$ does not have uniform
distribution although the individual chaotic signals and the noise
signals are uniform. This is because of the special way in which the
signals were added.

All the 24 chaotic signals and the noise signal were successfully
recovered in a lossless fashion at the receiver. We have used the
finite precision implementation of GLS-coding as proposed in
Appendix. This confirms the efficacy of Method 3.


\section{Remarks on the Three Methods} The following observations
can be made on the three methods:

\begin{enumerate}
\item The idea of symbolic sequence invariance is the key to the
success of all the three methods. The way this idea is implemented
is different in the three methods.

\item The way noise is handled is the same in Methods 1 and 2 since
there is no control on the noise in Scenario 1. Scenario 2 is much
more restrictive in terms of noise.

\item All the three methods rely on the finite precision implementation of GLS-coding, i.e.
finding the initial condition given an arbitrarily long symbolic
sequence of the chaotic signal (refer to Appendix). The method of
extracting the symbolic sequence from a ``noisy non-linear system''
and performing GLS-coding is analogous to filtering out noise in
linear systems by means of integration or other linear filters (for
eg., low pass filters).

\item Methods 2 and 3 can be easily extended to tent map, skew-tent
map, logistic map and other unimodal maps. It is also possible to
extend the methods to non-unimodal 1D maps and possibly higher
dimensional maps. The key is to find an analog of GLS-coding in
those cases, which we believe is possible.

\item Method 1 has a hint of using the idea of `forbidden
symbol'~\cite{Boyd1997, ARQ, GrangettoForbiddenSymbol, TrellisAC} by
allocating the length $q$ on the interval which is never used by the
map. This method can be potentially used for error correction and
detection.

\item In Method 2, just by observing the signal on the channel, no
information can be gleaned. The distribution is uniform and the
phase portrait is also random looking. The fact that multiple
chaotic signals have been embedded is not obvious. This property
enables it to be used in steganography or information hiding. In
LSB-steganography, the Least Significant Bit (LSB) of natural
signals is replaced by the secret (noise or noise-like). Method 2 is
doing the reverse: MSB-steganography, where the Most Significant Bit
(MSB) of the secret (noise or noise-like) is replaced by the
symbolic sequence of the chaotic signal.

\item Method 3 can handle any number of chaotic signals and one
noise signal. However, in practice there will be limitation on the
number of signals owing to finite precision since we are rescaling
the range of the signals to $[0,1)$ (by addition of 1 and division
by 3).

\item The methods show that chaotic signals are highly redundant and hence robust to noise. As
long as the symbolic sequence is preserved, the actual signal can be
distorted to a great deal. Also, forward iteration of chaotic
dynamical systems exhibits sensitive dependence on initial
conditions, but backward iteration shows resistance to noise. These
features are not exhibited by random/stochastic signals. This
probably makes a strong case for why biological systems may use
chaotic signals for transmission of information. Neuronal signals
may use similar mechanism for robust transport of information.

\item The above methods will not work for purely random signals or
for non-chaotic signals since there is no way we can construct the
entire trajectory by knowing the symbolic sequence. The redundancy
of chaotic signals is necessary. At the same time, chaotic signals
appear ``random'' in distribution.

\item In contrast with earlier work on multiplexing discrete chaotic
signals using chaotic synchronization, our methods can multiplex
more than two discrete chaotic signals. There are no coupling
coefficients used in our methods and hence there is no condition to
be satisfied for multiplexing. As noted previously, earlier methods
based on chaotic synchronization are vulnerable to noise and
parameter mismatch. Our methods are completely robust to any amount
of parameter mismatch since our methods do not rely on chaotic
synchronization. Methods 1 and 2 can tolerate noise but limited by
the number of signals that is added. However, in Method 3, the noise
magnitude is the same as that of the signals.

\item The methods that we have developed can be potentially used in
communication protocols, cryptography and steganography
applications.

\item There is no violation of Shannon's theorems for information
transmission in any of the methods. By transmitting the entire
trajectory, we are sending lots of bits, much more than actually
required for sending only the initial condition. These methods are
not meant for compression of data. These are mechanisms to exploit
the inherent redundancy in chaotic signals in spite of noise.

\end{enumerate}

\section{Conclusions and Open Problems}
 In this work, GLS-coding was used for multiplexing of
chaotic signals in the presence of noise. By using the idea of
symbolic sequence invariance, we were able to ``add'' several
chaotic signals and ``separate'' them losslessly at the receiver. We
can either have a lossy channel but with limited noise (Methods 1
and 2) or have a lossless channel with noise having the same
magnitude as the signal ``added'' in a very special way at the
sender (Method 3). An open problem is to investigate whether one can
have both features in a single method.

The inherent redundancy and structure in chaotic signals which
otherwise appear random in probability distribution can be harnessed
for robust communication of information. It is quite likely that
such efficient mechanisms (or similar ones) of handling noise in
dynamical systems are already being employed in naturally occurring
physical and biological systems.

Compared to existing methods of multiplexing discrete chaotic
signals, our methods are significantly superior in all respects. The
newly proposed methods can handle multiple signals from multiple
maps (including Bernoulli shift or the binary map which was not
possible by the method of Liu and Davis), completely robust to
parameter mismatch and good noise resistance capability.

\bibliographystyle{IEEEtran}
\bibliography{IEEEabrv,References}

\section*{Appendix: Finite Precision Implementation of GLS-coding} The idea of
GLS-coding~\cite{NithinGLS} is to first embed the stochastic
(binary) i.i.d source into the appropriate GLS and then treating the
message as a symbolic sequence, the initial condition is determined
by a backward iteration. This actually results in an interval (since
the symbolic sequence is finite in length) and the mid-point of the
interval is used as the initial condition. As the length of the
symbolic sequence increases, the interval in which the initial
condition is going to lie shrinks in size. This creates problem in
performing the backward iteration on a finite precision computer as
the two ends of the interval come very close to each other and at
some point it would be no longer possible to continue with the
backward iteration. This problem needs to be addressed by some kind
of re-normalization or re-scaling of the interval, in order for the
method to be useful for encoding long sequences. Another problem
with the method is that there is a long encoding delay. No output
can be written/sent unless the entire initial condition is
determined, which happens only after all the input bits are encoded.
Luca's chaotic compression method~\cite{Luca} also has similar
problems but is not addressed by them.

There has been efforts to address both these problems for Arithmetic
Coding~\cite{Langdon1981}. Since GLS-coding is an extension of
Arithmetic Coding, these methods could be used.

The idea is as follows. As soon as the interval completely lies to
the left of $p$ (for the standard tent map and binary map, $p=0.5$),
the final initial condition will have a `0' in its binary expansion
(`1' if the interval is completely to the right of $p$). Hence, this
can be written as output and the current interval can be doubled in
length. This ensures that the two ends of the interval will never
come close to each other. At the end of all the iterations, the
mid-point of the final interval is written as output. The only case
in which this method would fail is when the interval straddles 0.5
at every iteration. The probability of this happening exponentially
decreases with each iteration. To handle this special case, there
can be a check on the size of the interval and once it reduces to
certain value, the encoder is forced to generate an output and the
interval is reset to [0,1). The iteration starts afresh for the next
incoming bits. This would increase the size of the compressed file
slightly as we are not encoding the entire symbolic sequence to
determine a single initial condition, but the increase in size is
negligible for long sequences. A similar technique was used in the
IBM Q-coder~\cite{Langdon1981} to handle this problem. The algorithm
for encoding is described in Algorithm~\ref{alg:FinitePrecGLS}.  The
decoding algorithm is similar and is omitted here.

\begin{algorithm}[H]
\caption{Finite precision GLS-coding} \label{alg:FinitePrecGLS} %
\begin{enumerate}
\item Input: Binary message $M$ of length $N$ from stochastic binary i.i.d source $X$.
\item Compute probability of `0' as $p = \frac{total~~number~~of~~zeros~~in~~M}{N}$.
\item Embed source $X$ in to a GLS: construct GLS with partitions $[0,p)$ corresponding to symbol `0' and
partition $[p,1)$ corresponding to symbol `1'.
\item Initialize interval $[L, U)$ to $[0,1)$. Initialize $Tol =
10^{-8}$.
\item Initialize $k=1$.
\item Input the $k$-th bit from $M$.

\item If the bit is 0, then set:
\begin{eqnarray*}
L &\leftarrow& L/p\\
U &\leftarrow& U/p.
\end{eqnarray*}
else, set:
\begin{eqnarray*}
L &\leftarrow& (L-p)/(1-p)\\
U &\leftarrow& (U-p)/(1-p).
\end{eqnarray*}
Set $k \leftarrow k+1$.

\item if $0 \leq L,U < 0.5$, then Output bit `0' and set:
\begin{eqnarray*}
L &\leftarrow& 2L\\
U &\leftarrow& 2U.
\end{eqnarray*}
if $0.5 \leq L,U < 1$, then Output bit `1' and set:
\begin{eqnarray*}
L &\leftarrow& 2L-1\\
U &\leftarrow& 2U-1.
\end{eqnarray*}

\item if $ (U - L)   \leq Tol$, Output $x_{ic} = \frac{(L+U)}{2}$ in binary representation and reset
$[L,U) = [0,1)$.

\item If $k \leq N$, go to step 6, else continue to step 11.
\item If $L \leq 0.5 \leq U$, then $x_{ic} = 0.5$ else $x_{ic} =
\frac{(L+U)}{2}$.
\item Output $x_{ic}$ in binary representation.
\item Output $p$, $N$ and mode used as overheard information. The
precision of $p$ can be chosen conveniently.
\end{enumerate}
\end{algorithm}

For the multiplexing of chaotic discrete signals proposed in this
paper, we do not need to compute $p$ (in step 2 of
Algorithm~\ref{alg:FinitePrecGLS}), but can directly set it to
$p=0.5$ since we are using the standard binary map. The algorithm
described here can be extended easily to find the initial condition
from an arbitrarily long symbolic sequence of other chaotic maps
(for eg., to find the initial condition on the skew-tent map,
equation in step 7 needs to be modified appropriately). A similar
extension can be done for the Logistic map as well.

\end{document}